\documentclass[11pt]{article}
\usepackage{Blois,epsfig}
\usepackage{graphics}
\usepackage{float}
\def\be{\begin{equation}}
\def\ee{\end{equation}}
\def\bea{\begin{eqnarray}}
\def\eea{\end{eqnarray}}

\begin{document}
\vspace*{2cm}
\title{TOTEM PHYSICS}

\author{G. Anelli$^{1}$, A. Aurola$^{2}$, V. Avati$^{1}$,  V. Berardi$^{3}$,  
U. Bottigli$^{4}$, M. Bozzo$^{5}$, 
E. Br\"{u}cken$^{2}$, A. Buzzo$^{5}$, M.~Calicchio$^{3}$, F. Capurro$^{5}$, 
M.G. Catanesi$^{3}$, M.A. Ciocci$^{4}$, S. Cuneo$^{5}$, 
C. Da Vi\`{a}$^{6}$, M. Deile$^{1}$, E.~Dimovasili$^{1}$, 
K. Eggert$^{1}$, M. Er\"{a}luoto$^{2}$, F. Ferro$^{5}$, A. Giachero$^{5}$, 
J. Hasi$^{6}$, F. Haug$^{1}$, J. Heino$^{2}$, T.~Hilden$^{2}$, 
P. Jarron$^{1}$, J.~Kalliopuska$^{2}$, J. Ka\v{s}par$^{7}$,
A. Kok$^{6}$, V. Kundr\'{a}t$^{7}$, K. Kurvinen$^{2}$, 
S. Lami$^{4}$, J.~L\"{a}ms\"{a}$^{2}$, G. Latino$^{4}$, R. Lauhakangas$^{2}$, 
E. Lippmaa$^{8}$, 
J. Lippmaa$^{2}$, M. Lokaj\'{\i}\v{c}ek$^{7}$, M. LoVetere$^{5}$, 
D.~Macina$^{1}$, 
M.~Macr\'{\i}$^{5}$, M. Meucci$^{4}$, S.~Minutoli$^{5}$, A. Morelli$^{5}$, 
P. Musico$^{5}$, M. Negri$^{5}$, H.~Niewiadomski$^{1}$, E.~Noschis$^{1}$, 
J. Ojala$^{2}$, F. Oljemark$^{2}$, R. Orava$^{2}$, M. Oriunno$^{1}$, 
K. \"{O}sterberg$^{2}$, R.~Paoletti$^{4}$, 
A.-L.~Perrot$^{1}$, 
E.~Radermacher$^{1}$, E. Radicioni$^{3}$, E. Robutti$^{5}$, 
L. Ropelewski$^{1}$, G.~Ruggiero$^{1}$, 
A. Rummel$^{8}$,
H.~Saarikko$^{2}$, G. Sanguinetti$^{4}$,
A. Santroni$^{5}$, S. Saramad$^{1}$, F. Sauli$^{1}$, A.~Scribano$^{4}$, 
G. Sette$^{5}$, J.~Smotlacha$^{7}$, W. Snoeys$^{1}$, C. Taylor$^{9}$, 
A. Toppinen$^{2}$, A. Trummal$^{8}$,
N.~Turini$^{4}$, N. Van Remortel$^{2}$, L. Verardo$^{5}$, 
A.~Verdier$^{1}$, S.~Watts$^{6}$, J. Whitmore$^{10}$}

\address{~~\\$^{1}$CERN, Gen\`{e}ve, Switzerland,\\ 
$^{2}$Helsinki Institute of Physics and University of Helsinki, Finland,\\
$^{3}$INFN Sezione di Bari and Politecnico di Bari, Bari, Italy,\\  
$^{4}$Universit\`{a} di Siena and Sezione INFN-Pisa, Italy,\\ 
$^{5}$Universit\`{a} di Genova and Sezione INFN, Genova, Italy,\\ 
$^{6}$Brunel University, Uxbridge, UK,\\ 
$^{7}$Academy of Sciences of the Czech Republic and Institute of Physics, 
Praha, Czech Republic,\\ 
$^{8}$National Institute of Chemical Physics and Biophysics NICPB, Tallinn,
Estonia.\\
$^{9}$Case Western Reserve University, Dept. of Physics, Cleveland, OH, USA,\\
$^{10}$Penn State University, Dept. of Physics, University Park, PA, USA\\
\vspace*{5mm}
Presented by K. Eggert}

\maketitle
\abstracts{
This article discusses the physics programme of the TOTEM experiment at the 
LHC. A new special beam optics with $\beta^{*} = 90$\,m, enabling the 
measurements of the total cross-section, elastic pp scattering and 
diffractive phenomena already at early LHC runs, is explained. 
For this and the various other TOTEM running scenarios,
the acceptances of the leading proton detectors and of the forward 
tracking stations for some physics processes are described.
}

\newpage
\section{Introduction}

The physics programme of the TOTEM experiment can be performed in several 
few-days runs per year with special optics during the first three years of 
the ``Large Hadron collider'' (LHC) operation. As stated in the 
Letter of Intent~\cite{totem_loi} and the Technical Design 
Report~\cite{totem_tdr}, the principal goals of TOTEM are:
\begin{itemize}
\item the measurement of the total cross-section with a precision of about 
1\,mb using, via the Optical Theorem, the luminosity independent method that 
requires the simultaneous measurements of the total inelastic rate and the 
elastic pp scattering down to four-momentum transfers ($-t \approx p^{2} 
\theta^{2}$) of a few $10^{-3}\,\rm GeV^{2}$;
\item the measurement of the elastic pp scattering over a wide range in $-t$, 
up to $\rm 10\,GeV^{2}$; and
\item the study of inelastic diffractive final states that will comprise 
almost a quarter of all interactions.
\end{itemize}
The TOTEM programme focuses on physics not available to the general purpose 
experiments at the LHC. The experimental set-up, as described in the 
TDR~\cite{totem_tdr} and in these proceedings~\cite{blois_gennaro}, 
has therefore to be capable of meeting the challenges of detecting 
inelastically produced particles in the very forward region as well as 
protons very close to the LHC beams. It is comprised of Roman Pot detectors 
for the leading proton measurement together with tracking stations T1 and T2 
in the very forward region of CMS. 
\begin{figure}[h!]
\begin{center}
\mbox{
\epsfig{file=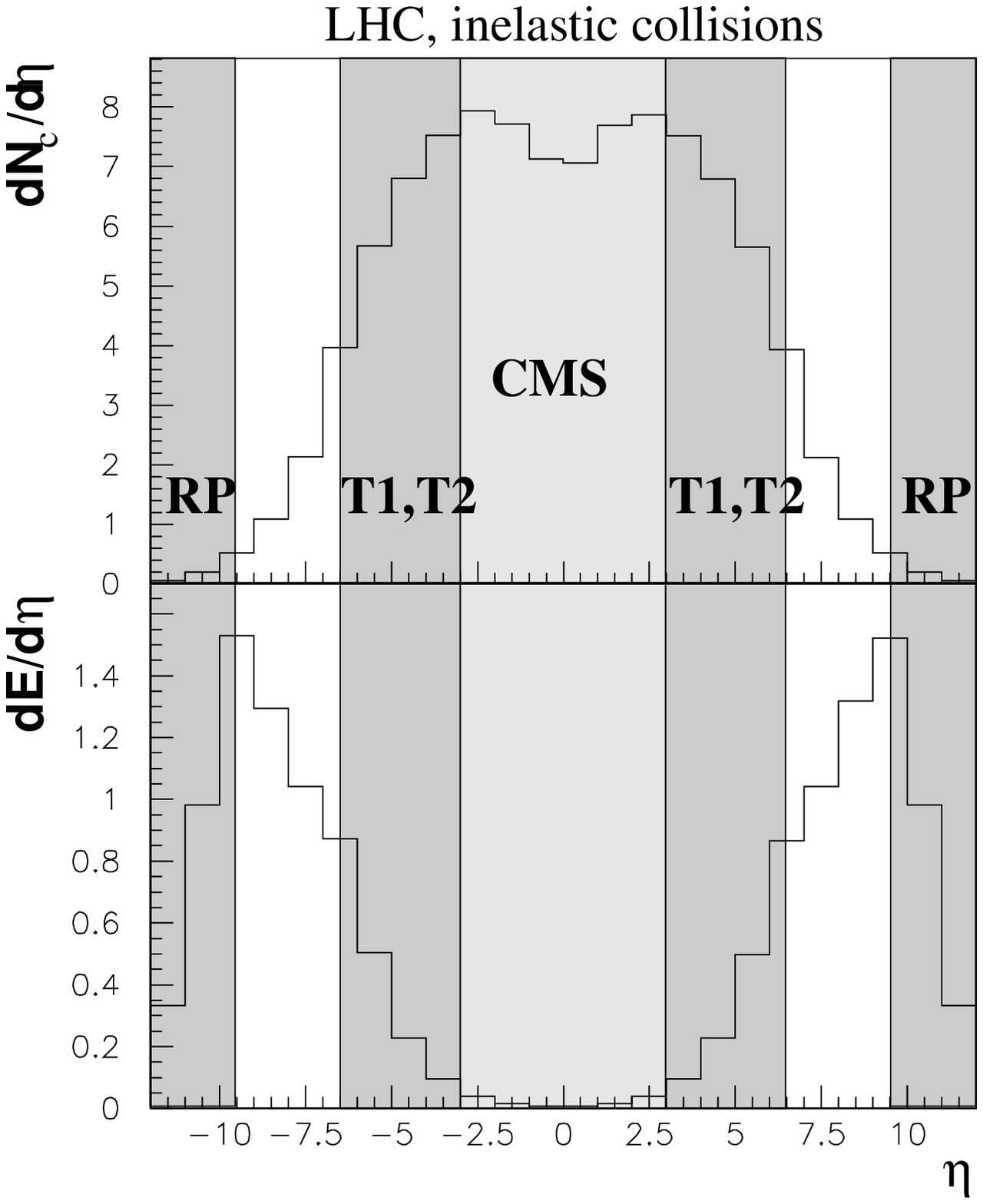,height=7cm}
\hspace*{5mm}
\epsfig{file=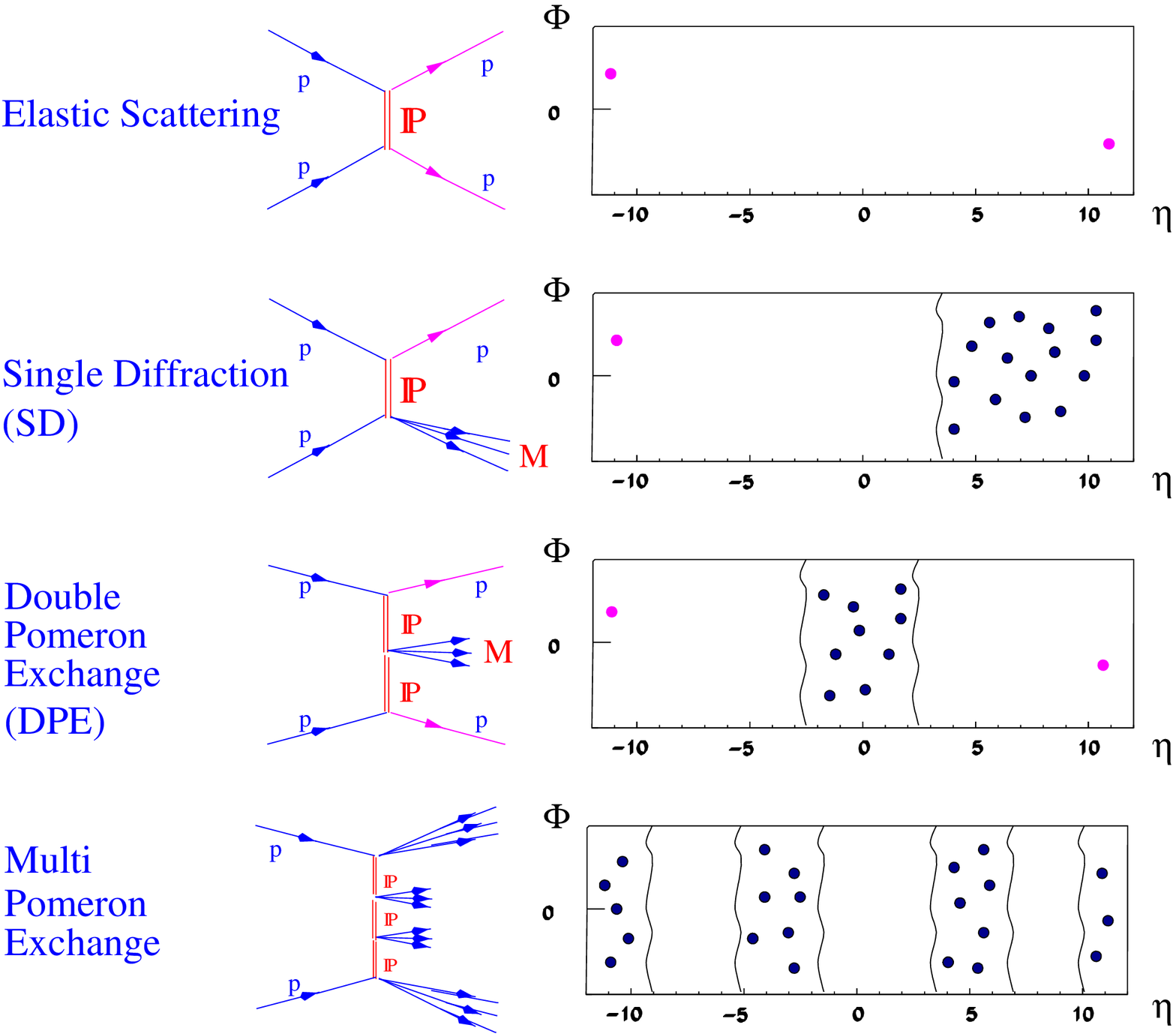,height=7cm}
}
\caption{Left: pseudorapidity distribution of the charged particle multiplicity
(upper part) and of the energy flow (lower part) per generic inelastic event; 
the grey bands correspond to the acceptance of the TOTEM detectors, while CMS
covers the centre.
Right: some diffractive process classes as Feynman diagrams, and 
their particle distributions in the rapidity-azimuth plane. 
}
\label{fig_1}
\end{center}
\end{figure}

The combined CMS and TOTEM detectors cover an unprecedented pseudorapidity 
range for charged and neutral particles together with an excellent acceptance 
for leading protons. This is demonstrated in Fig.~\ref{fig_1} (left) 
where the charged particle multiplicity and the energy flow in generic 
inelastic events is shown together with the common CMS/TOTEM acceptance. 
It is obvious that detectors like ATLAS and CMS, being optimised for hard QCD 
processes, will miss most of the energy. Never before has an experiment been 
constructed at a hadron collider with such a wide 
acceptance together with an excellent measurement of the leading protons.  
The large coverage of such a ``combined experiment'' will allow for unique 
measurements that will be discussed in 
the near future in a common CMS/TOTEM physics report. The wide variety of 
diffractive processes that can be addressed is sketched in the 
rapidity-azimuth distributions (Fig.~\ref{fig_1}, right) taken from the FELIX 
proposal~\cite{felix}. Starting with the cleanest process of elastic 
scattering up to the most complicated structure of  multi-gap events due 
to multi-Pomeron exchange, diffraction can be precisely studied from soft 
to very hard processes, including high-$p_{T}$ jets, heavy particles, W and Z, 
and possibly new particles, as discussed in Ref.~\cite{felix_jphys}. 
As an example, Double 
Pomeron Exchange (DPE) offers the possibility of turning the LHC into an 
almost clean gluon factory, constrained by the 
measured momentum loss of the two forward protons. In this way, precise 
QCD studies and a search for new particles, created by gluon-gluon 
interactions, can be performed.

Furthermore, the observation of events over the full range of pseudorapidity 
can contribute significantly to cosmic-ray astrophysics and to the 
resolution of some of its outstanding problems~\cite{astro}.  
It will open previously inaccessible regions of parameter space to 
quantitative studies where cosmic ray experiments have hinted at unusual 
physics.

\section{Running Scenarios}
\subsection{Overview}
The versatile programme of TOTEM requires different running scenarios that 
have to be adapted to the LHC commissioning in the first few years. 
A summary is given in Tab.~\ref{tab_scenarios}. 
\begin{table}[h!]
\caption{
TOTEM(+CMS) Diffractive Running Scenarios ($k$: number of bunches,
$N$: number of protons per bunch)
\label{tab_scenarios}}
\begin{tabular}{|c|cccccl|}
\hline
Scen. & $\beta^{*}$[m] & $k$ & $N/10^{11}$   & $\mathcal{L} [{\rm cm^{-2}s^{-1}}]$ &
$|t|$-range [GeV$^{2}$] & Objectives\\
\hline
1        & 1540        & 43   & 0.3          & $2\times10^{28}$ & 
0.002 $\div$ 1.5 & low $|t|$ elastic, $\sigma_{T}$, \\
         &             &      &              &                  & 
               & min. bias, soft diffraction\\
2        & 1540        & 156  & 0.6$\div$1.15 & $2\times10^{29}$ &
0.002 $\div$ 1.5 & diffraction \\
3        &   90        & 156  & 1.15         & $3\times10^{30}$ &
0.03 $\div\, \sim$2   & intermediate $|t|$ elastic, $\sigma_{T}$\\
4        &   90        & 936  & 1.15         & $2\times10^{31}$ &
0.03 $\div\, \sim$2   & (semi-) hard diffraction\\
5        &   18        & 2808 & 1.15         & $3.6\times10^{32}$ &
0.6 $\div$ 8   & large $|t|$ elastic \\
6        &   0.5       & 2808 & 0.3          & $10^{33}$ &
1 $\div\, \sim$10     & rare diffractive processes\\
\hline
\end{tabular}
\end{table}

Contrary to other detectors, the leading proton measurement critically 
depends on the focussing of the beams at the Intersection Point (IP), 
expressed by the betatron value ($\beta^{*}$). For the low-$t$ measurements, 
the detection of protons with scattering angles of a few microradians 
requires the development of special beam optics with a $\beta^{*}$ value as 
large as possible. 
The dedicated TOTEM optics with $\beta^{*} = 1540\,$m are characterised by 
a small beam divergence (0.3\,$\mu$rad) and by parallel-to-point focussing 
in both projections at the location of the Roman Pot detectors, 220\,m away 
from the IP. As a consequence, the beam size at the IP is large, and 
at most 156 bunches are allowed in the machine to avoid parasitic bunch 
crossings, thus leading to low luminosities. 

At the start of the LHC, the machine will be operated with only a few bunches 
and with a reduced proton density per bunch and zero degree crossing angle. 
The beams will then be gradually squeezed to $\beta^{*}$ values around a 
few metres. TOTEM might request at that time a relaxation of the beams to 
$\beta^{*} = 90\,$m for some few-days runs. This would make a 
measurement of elastic scattering with $-t$ values above 0.03\,GeV$^{2}$ 
possible and allow an early crude determination of the total cross-section. 
As discussed later, this optics scenario would also 
provide an excellent measurement of the momentum loss of diffractive protons, 
opening the studies of soft and semi-hard diffraction.

With the nominal optics ($\beta^{*} = 0.5\,$m), TOTEM will register 
elastically scattered protons at large $|t|$ values and diffractive protons 
with a momentum loss above 2\,\%. In this running condition, TOTEM and CMS 
together can study hard diffraction. 

\subsection{Optics with an Intermediate $\mathbf{\beta^{*}}$}

\begin{figure}[H]
\begin{center}
\mbox{
\psfig{file=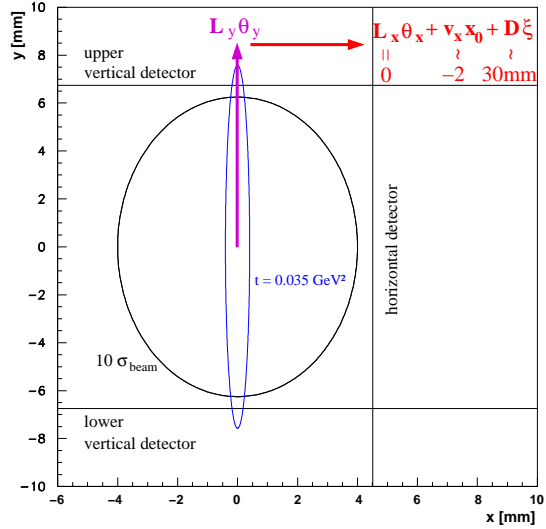,height=7cm}
}
\end{center}
\caption{Principle of measuring the momentum loss $\xi$ of diffractively 
scattered protons with the $\beta^{*} = 90\,$m optics. The large  
effective length $L_{y}$ acts on the protons' vertical scattering angle 
$\theta_{y}$ to move them out of the beam, whereas the vanishing $L_{x}$ 
maximises the sensitivity of the horizontal proton position to $\xi$.
}
\label{fig_90mplot}   
\end{figure}
The proton acceptance and measurement accuracy depend critically  
on the machine optics. While the ultimate optics for lowest $t$ with a 
$\beta^{*} = 1540\,$m needs a dedicated injection, complicating the 
commissioning, an intermediate optics with a standard injection 
might be more easily achievable at the LHC start. We consider an optics with a 
$\beta^{*}$ of 90\,m which can be obtained by unsqueezing the beams after 
injection. 

The basic considerations of this optics are illustrated in 
Fig.~\ref{fig_90mplot}. At the Roman Pot 
position, the deflection of a proton produced in a pp interaction at the IP 
with transverse coordinates $(x_{0}, y_{0})$, a momentum loss 
$\xi = \Delta p / p$ and a polar angle $\theta \approx \sqrt{-t}/p$ can be 
generally expressed by the machine parameters, using the effective length $L$ 
and the magnification $v$:  
\begin{eqnarray}
x &=& L_{x}\, \theta_{x} + v_{x}\, x_{0} + D \, \xi \\
y &=& L_{y}\, \theta_{y} + v_{y}\, y_{0} 
\end{eqnarray}
To push the protons vertically into the acceptance of the forward detectors, 
the optics was optimised towards a large $L_{y}$ and $v_{y} = 0$ 
(vertical parallel-to-point focussing). With $L_{y} = 270\,$m, 
protons with $-t > 3 \times 10^{-2}\,\rm GeV^{2}$ start to be observed in the 
detectors. The horizontal deflection depends on the momentum loss $\xi$,
the horizontal production angle $\theta_{x}$, and the transverse coordinate 
$x_{0}$ at the IP. 
By choosing $L_{x} = 0$, the emission angle dependence is 
eliminated. If the vertex coordinate at the IP can be measured with a 
precision of about 10\,$\mu$m by the inner tracker of CMS, the momentum loss 
of diffractive protons can be measured with a precision $\sigma(\xi)$ 
better than $10^{-3}$. In the case of elastically scattered protons, 
the dependence on the vertex position drops out due to the requirement of 
collinearity between the two protons.

In summary, the above optics could probably be commissioned at the beginning 
with not too many complications. It would allow the first measurements of the 
total cross-section and elastic scattering at low to intermediate $t$-values. 
The acceptance for diffractive protons is about 60\,\%, due to a good 
acceptance at low $t$-values, independent of the proton momentum loss.

\section{Physics Goals of TOTEM}
\begin{figure}[h!]
\begin{center}
\mbox{
\epsfig{file=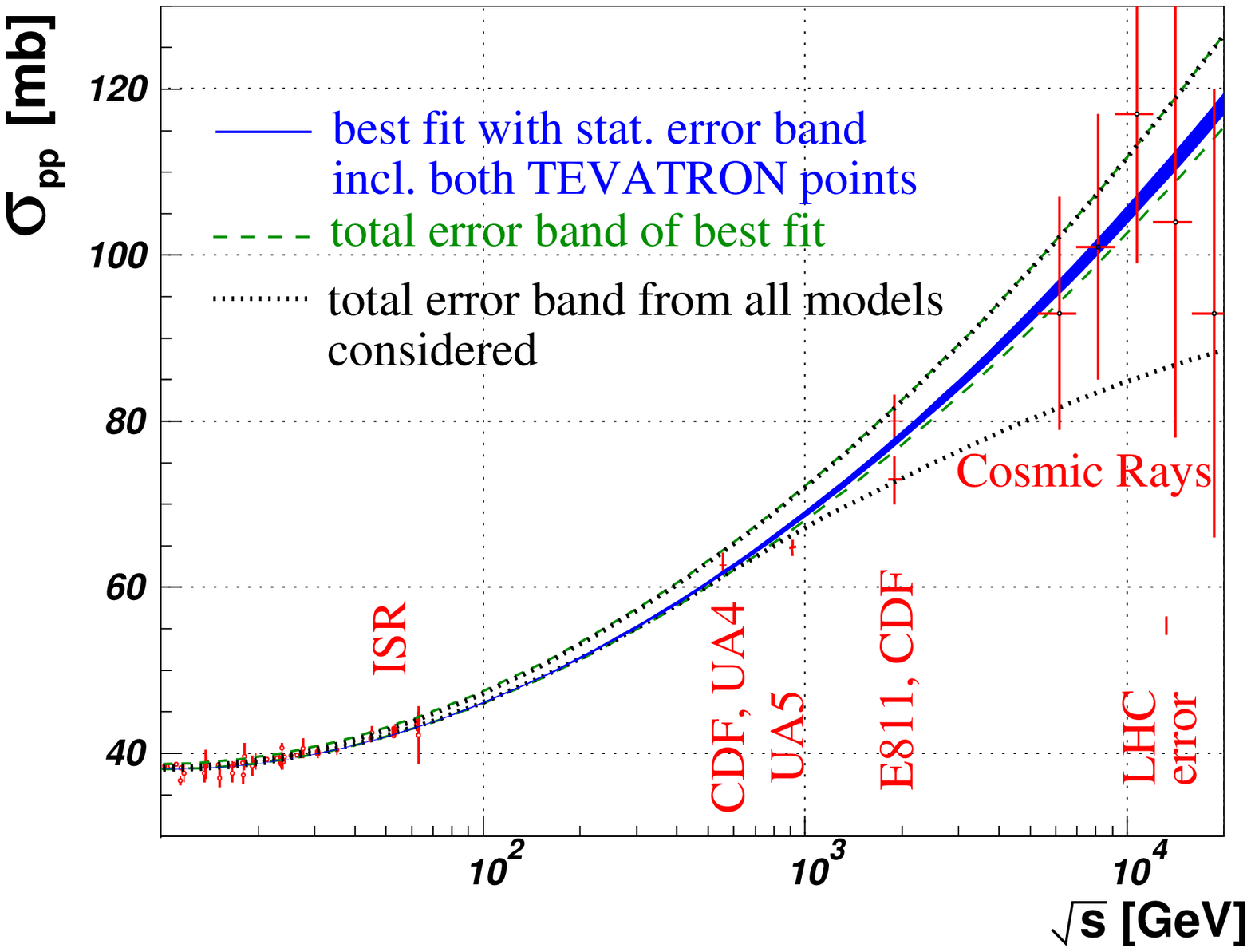,height=6cm}
\epsfig{file=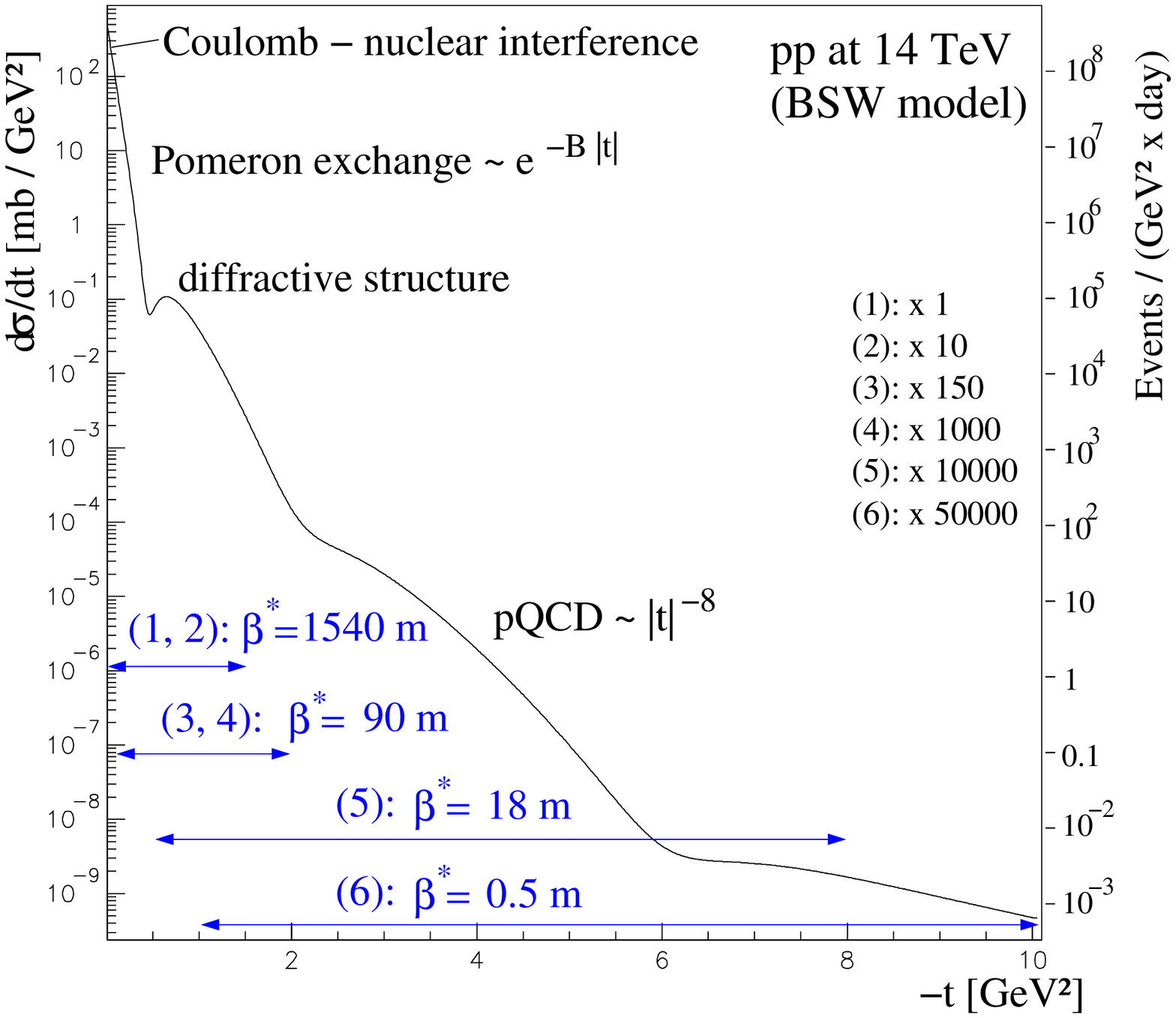,height=7cm}
}
\end{center}
\caption{Left: COMPETE fits~\protect\cite{compete} to all available $pp$ and 
$p\bar{p}$ 
scattering data with statistical (blue solid) and total (dashed) error 
bands, the
latter taking into account the Tevatron ambiguity. The outermost curves 
(dotted) give the total error band from all parameterisations considered.
Right: prediction for elastic $pp$ scattering at LHC with 
one-day statistics for the running 
scenarios defined in Tab.~\ref{tab_scenarios}.
}
\label{fig_sigmatot}   
\end{figure}

Without any doubt, a precise measurement of the pp total cross-section 
$\sigma_{tot}$ (Fig.~\ref{fig_sigmatot}, left) and of the elastic scattering 
over a large $t$-range (Fig.~\ref{fig_sigmatot}, right) is of 
primary importance in distinguishing between 
different models of soft proton interactions which exhibit significantly 
different cross-sections at large energies and large $t$-values. 

\subsection{Total Cross-Section}
Fig.~\ref{fig_sigmatot} (left) summarises the existing data from low energies 
up to collider and cosmic-ray energies. Unfortunately, cosmic-ray data have 
large uncertainties, and the conflicting data from the TEVATRON demonstrate 
the difficulty of these measurements and make an extrapolation to 
higher energies uncertain, leaving a wide range for the expected value of 
the total cross-section at the LHC energy of $\sqrt{s} = 14\,\rm TeV$, 
typically from 90 to 130\,mb. Taking into account all available data, 
the COMPETE collaboration~\cite{compete} has made an overall fit of the 
energy dependence of the total cross-section and the ratio $\rho$ of the real 
to imaginary parts of the elastic scattering amplitude, and predict for the 
LHC:
\begin{equation}
\sigma_{tot} = 111.5 \pm 1.2^{+4.1}_{-2.1}\,{\rm mb}\:,\quad 
\rho = 0.1361\pm0.0015^{+0.0058}_{-0.0025}\:.
\end{equation}
The precision of the extrapolation is dominated by the ambiguity in the 
TEVATRON data (second error).

The total pp cross-section is related to the nuclear elastic forward 
scattering amplitude via 
the Optical Theorem. This allows a luminosity independent determination 
based on the total inelastic rate and the extrapolation of the pp elastic 
scattering to the optical point $t = 0$. Hence a simultaneous measurement 
of the total inelastic rate and the elastic scattering at the lowest 
possible $|t|$-values, which requires a large $\beta^{*}$, is needed.

\begin{table}[h!]
\begin{center}
\begin{tabular}{|l|c|c|c|c|}
\hline
Process      &  $\sigma$ [mb] & \multicolumn{3}{c}{Trigger Losses [mb]}\\ 
             &                & Double Arm & Single Arm & After Extrapol. \\
\hline
Minimum bias       & 58       & 0.3        &  0.06      & 0.06 \\
Single Diffractive & 14       & --         &  2.5       & 0.6  \\
Double Diffraction &  7       & 2.8        &  0.3       & 0.1  \\ 
Double Pomeron     &  1       & --         &  --        & 0.02 \\
Elastic Scattering & 30       & --         &  --        & 0.1  \\ 
\hline
\end{tabular}
\caption{Trigger losses for double-arm, single-arm and proton trigger (the 
latter for Double Pomeron Exchange and elastic scattering).
In the double-arm trigger at least one particle per hemisphere has to be
detected, while in the single-arm only at least one particle in either
hemisphere is required.} 
\label{tab_inelastic}
\end{center}
\end{table}
To determine the inelastic rate, TOTEM has made extensive trigger studies 
that are summarised in Tab.~\ref{tab_inelastic}. Algorithms have been 
developed to trigger on charged particles in the two large-acceptance  
forward telescopes and in the Roman Pot detectors (see Fig.~\ref{fig_1}, left).
Inelastic events are recorded either with a double arm trigger or with a 
single arm trigger that contains more background mainly from beam-gas events. 
While most of the non-diffractive events are registered, the largest trigger 
losses occur in single diffraction. Single diffractive events with 
masses below 10\,GeV fail the trigger (see the mass acceptance curve in 
Fig.~\ref{fig_massaccept}, left), resulting in a trigger loss of 2.5\,mb 
which, however, can be corrected for, using 
$\frac{d\sigma}{dM} \propto \frac{1}{M}$. 
Triggers on just beam crossings will help in 
the understanding of the trigger algorithms and of the background. 
The final uncertainties in the inelastic rate are estimated to be around 
0.8\,mb.

\begin{figure}[h!]
\vspace*{-5mm}
\begin{center}
\mbox{
\psfig{file=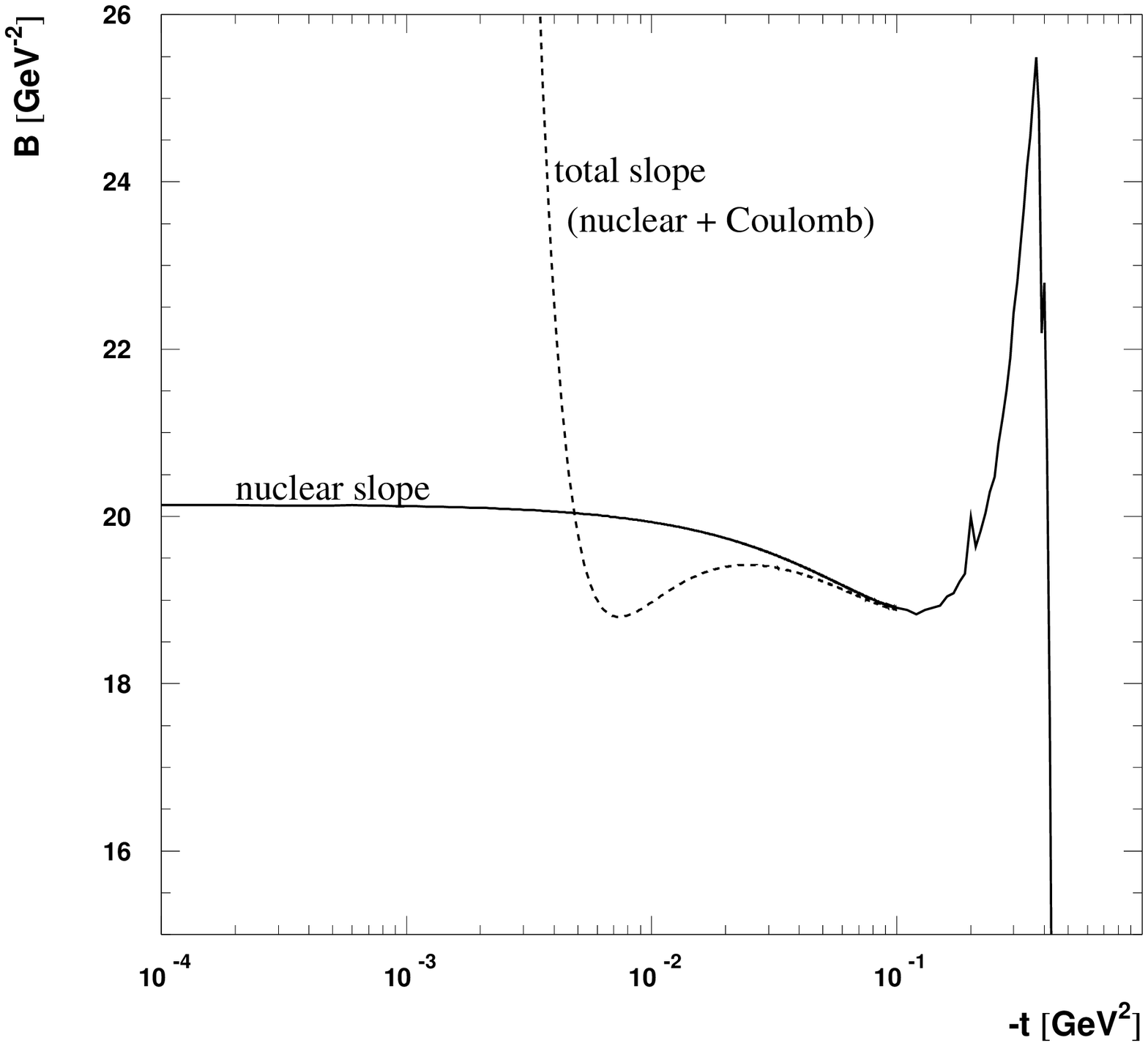,height=7cm}
\hspace*{5mm}
\psfig{file=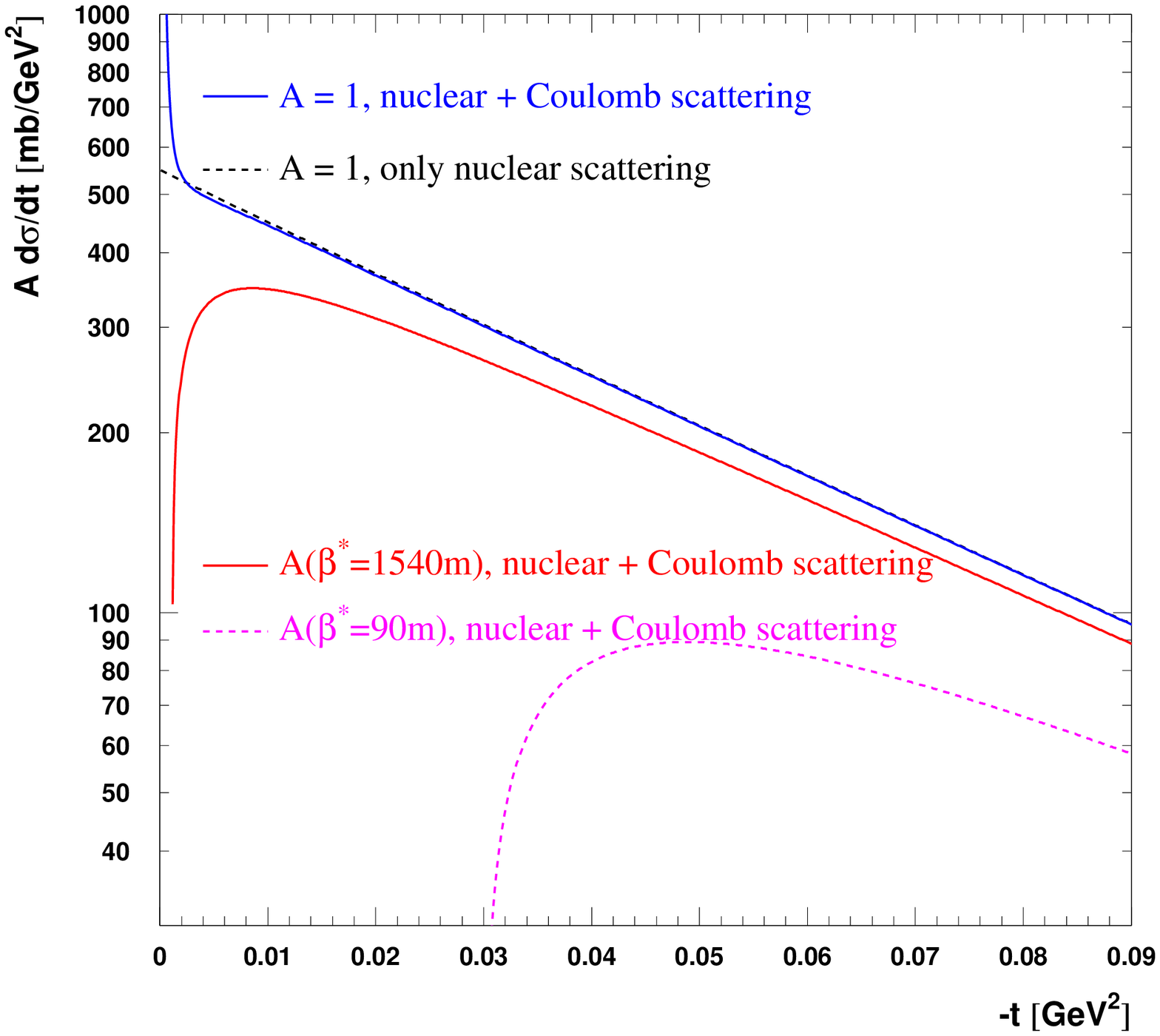,height=7cm}
}
\end{center}
\caption{Left: slope parameter 
$B(t) \equiv 
-\frac{\frac{d}{dt}\left(\frac{d\sigma}{dt}\right)}{\frac{d\sigma}{dt}}$ 
of elastic pp scattering with and without Coulomb contribution. Right:
differential cross-section of elastic scattering without (top curves) and
with (bottom curves) acceptance effect from the Roman Pots at high and 
intermediate $\beta^{*}$.
}
\label{fig_extrapol}   
\end{figure}
The elastic pp scattering with a cross-section of about 30\,mb has 
to be measured to the lowest possible $|t|$-values to enable an extrapolation 
to the optical point $t = 0$. The relative statistical uncertainty of the 
extrapolation is estimated to be 0.1\,\% based on 10 hours of data taking 
with the large $\beta^{*}$ optics. Systematic effects, such as the alignment
of the detectors with respect to the beams, have to be studied carefully and
will probably dominate the error of the extrapolation.
Furthermore, this extrapolation is also model 
dependent due to the way the nuclear--Coulomb interference is treated which 
influences the exponential slope of the nuclear scattering up to a few 
$\rm 10^{-2}\,GeV^{2}$. The slope parameter $B$ of the expected 
exponential distribution is given in Fig.~\ref{fig_extrapol} (left) 
with and without 
Coulomb interference. Coulomb effects become negligible for 
$|t| > 2\times 10^{-2}\,\rm GeV^{2}$. Fig.~\ref{fig_extrapol} (right) shows 
the elastic scattering distribution, also multiplied by the acceptances of 
the 1540\,m and 90\,m optics. 
Taking the theoretical uncertainties into 
account, an ultimate extrapolation precision of about 0.5\,\% can be 
achieved with the highest-$\beta^{*}$ optics. 
With the acceptance down to $|t| \sim 3 \times 10^{-3}\,\rm GeV^{2}$,
indications of the Coulomb term will become visible.  

It is worth mentioning that, under the assumption of an early commissioning 
of an optics with $\beta^{*} = 90$\,m, the exponential behaviour of the 
$t$-distribution can be measured in the range 
$0.04\,{\rm GeV}^{2} < -t < 0.2\,{\rm GeV}^{2}$, 
allowing the determination of the rate at $t = 0$ with a precision 
better than 5\,\%.  The total cross-section and the luminosity would then 
be known at early data taking with a similar accuracy. In addition, the 
transverse vertex distribution at the interaction point ($\sim$\,0.2\,mm) 
could be determined with elastic events which would lead to another 
determination of the luminosity, using the machine parameters:
\begin{equation}
\mathcal{L} = \frac{N^{2}\,k\,f}{4\,\pi\,\sigma_{x}\,\sigma_{y}} = 
\frac{I^{2}}{4\,\pi\,f\,k\,\sigma_{x}\,\sigma_{y}}
\end{equation}

\subsection{Elastic pp Scattering}
Much of the interest in large-impact-parameter collisions centres on elastic 
scattering and soft inelastic diffraction. High-energy elastic nucleon 
scattering represents the collision process in which the most precise data 
over a large energy range have been gathered. The differential 
cross-section of elastic pp interactions at the LHC, as predicted by the BSW 
model~\cite{bsw}, is given in Fig.~\ref{fig_sigmatot} (right). 
Increasing the squared momentum 
transfer, $-t$, means looking deeper into the proton at smaller distances. 
Several $t$-regions with different scattering behaviours can be identified. 
For $|t| < 10^{-3}\,\rm GeV^{2}$, Coulomb scattering ($\propto |t|^{2}$) is 
dominant,
whereas for $|t| > 10^{-3}\,\rm GeV^{2}$, nuclear scattering via Pomeron 
exchange takes over ($\propto \exp(-B |t|)$), 
with nuclear-Coulomb interference in between, thus allowing a measurement 
of the $\rho$ value. 
At large $t$-values above 1\,GeV$^{2}$, perturbative QCD with e.g. 
triple-gluon exchange ($\propto |t|^{-8}$) might 
describe the central elastic collisions of the proton. It is obvious that 
many different models try to describe the behaviour of the elastic scattering. 
In particular, the regime of large spacelike $|t|$ is associated with small 
interquark transverse distances within a proton. Large differences between 
the models are expected, and hence a high-precision measurement up to 
$|t| \approx 10\,\rm GeV^{2}$ will help to better understand the structure 
of the proton.

The elastic scattering distribution extends over 11 
orders of magnitude and has therefore to be measured with 
several different optics scenarios, as described in Tab.~\ref{tab_scenarios}. 
Decreasing $\beta^{*}$ shifts the observable $|t|$-range to larger values and 
simultaneously increases the luminosity, compensating the drastic decrease 
of the cross-section. Even at the largest accepted $|t|$-values 
($\sim 10\,\rm GeV^{2}$), limited by the size of the beam tube, 
about 100 events\,/\,GeV$^{2}$ are expected for a one-day run. 
Double Pomeron Exchange will be a substantial background to elastic 
scattering since the 
$t$-distribution is much flatter. A careful check on the collinearity of the 
two protons is therefore mandatory.

\begin{figure}[h!]
\vspace*{-1mm}
\begin{center}
\mbox{
\epsfig{file=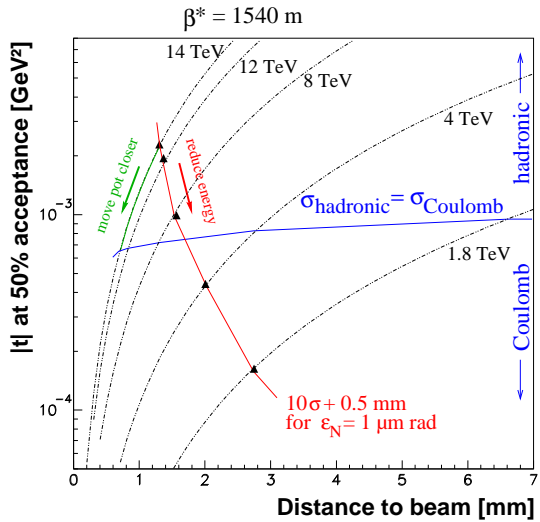,height=7cm}
}
\end{center}
\vspace*{-3mm}
\caption{$t$-value where 50\,\% acceptance is reached, as a function of the 
detector distance from the beam centre, for different centre-of-mass 
energies. The nominal TOTEM operating point is at 14\,TeV and a normalised 
emittance of 1\,$\mu$m\,rad implying 
a detector-beam distance of 10\,$\sigma + 0.5$\,mm = 1.3\,mm. 
To access the Coulomb region, either the distance between
detector and beam or the energy has to be reduced.}
\label{fig_rho}   
\vspace*{-2mm}
\end{figure}

The measurement of the Coulomb/nuclear interference and hence of the $\rho$ 
parameter is best done when Coulomb and nuclear scattering are of the same 
order. Demonstrating the possibilities of a rho measurement, Fig.~\ref{fig_rho}
displays the $t$-value at 50\,\% detection efficiency versus 
the detector distance to the beam for different collider energies. 
As the closest approach to the beam, 10 times the rms beam spread plus 0.5\,mm 
is considered (line with triangles). 
Thus, at the nominal LHC energy with normal beams, the Coulomb region is not 
reachable. The almost horizontal line indicates the t-values at which the 
cross-sections for hadronic and Coulomb scattering are equal. 
The two ways to approach the Coulomb region are indicated by the arrows. 
For very low beam intensities and consequently thin beams (low beam 
emittances), it might be possible to approach the beams closer. A safer 
way is to reduce the LHC energy. The larger beam spread at lower energies 
($\sigma(\theta)\propto s^{-\frac{1}{4}}$) is largely compensated by the 
lower $|t|$-values at the 
same emission angle ($-t \approx s\, \theta^{2}$). 
Hence a $\rho$ measurement at an energy of 8\,TeV should be possible.

\subsection{Diffractive Physics}
Most of the soft inelastic diffraction processes via single or double 
Pomeron exchange (see Fig.~\ref{fig_1}, right) are peripheral, occurring at 
collision impact parameters around 1.5\,fm. The proton(s) involved may stay 
intact or dissociate. The events are characterised by the momentum 
loss $\xi$, the squared four-momentum transfer $t$ and the azimuthal angle 
$\phi$ of the proton and/or by a rapidity gap $\Delta \eta$
which is related to $\xi$ ($\Delta \eta = -\ln \xi$). 
At the start of the LHC, when the luminosities are low, TOTEM will concentrate 
on measurements of the individual diffractive cross-sections -- single and 
double diffraction, Double Pomeron Exchange (DPE) -- and their dependence on
$\xi$ and $t$. Furthermore, DPE processes leading to low mass final states are 
an important potential resource for spectroscopy.

With increasing luminosity, semi-hard and hard diffraction -- i.e. diffractive 
processes that contain a hard (short-distance) collision with visible jets 
in the final state -- will come into our reach. 
The trigger will still be on the forward protons detected in the Roman Pots
or on dissociated protons with forward particles in T1 or T2. The basic hard 
diffractive processes with one or more gaps in the final state are linked 
together. To really understand these events, one will need to examine 
the $t$-distribution of the forward protons for different hardness of the 
collision process (jet transverse momenta), as well as to study the 
generalisation to the cases where the proton undergoes soft diffraction. 
Correlating the jet activities and the jet transverse momenta with the 
parameters of the forward protons will give a first glance to the 
understanding of the underlying dynamics. Extraordinarily clean events are 
those with only two coplanar jets and two unfragmented protons. 
It is obvious that for these physics a combined CMS/TOTEM detector is needed.
\begin{figure}[h!]
\begin{center}
\mbox{
\epsfig{file=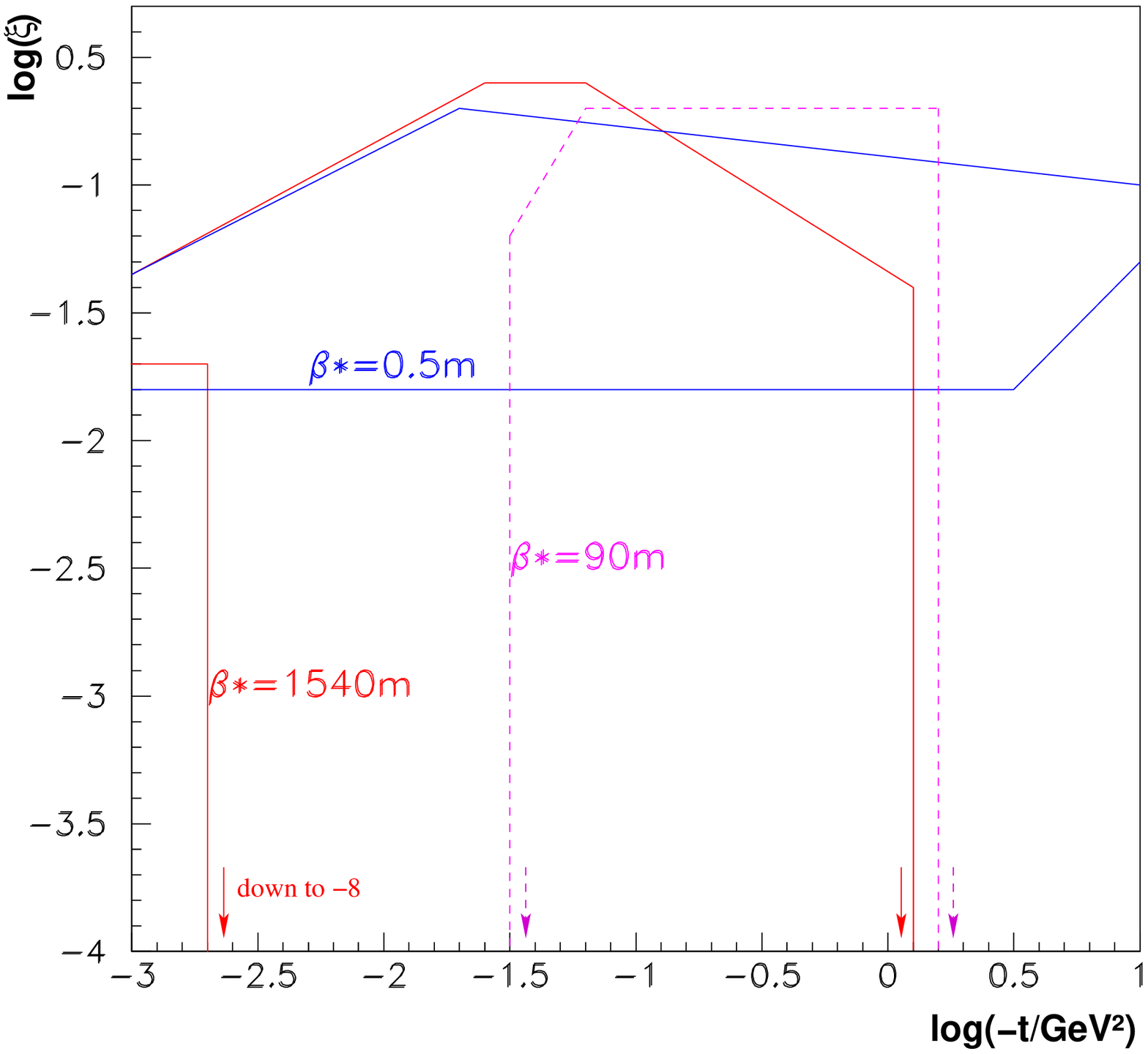,height=7cm}
\epsfig{file=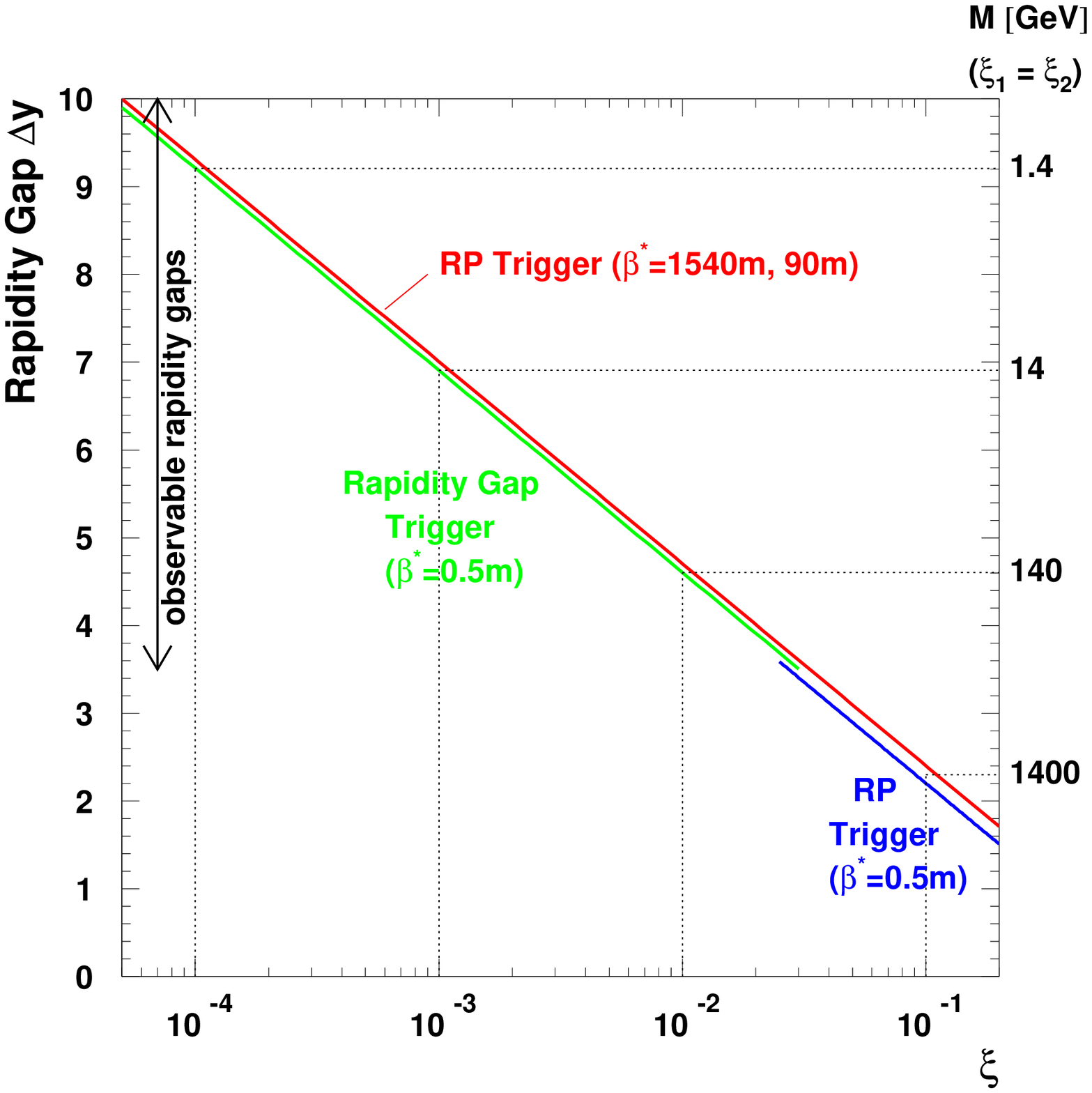,height=7.5cm}
}
\end{center}
\caption{Left: acceptance of the RP station at 220\,m for leading protons with 
squared four-momentum transfer $t$ and momentum loss $\xi$. For each of the 
three optics scenarios with $\beta^{*} =$ 1540\,m, 90\,m and 0.5\,m the contour
lines of the 50\,\% acceptance level are drawn. Right: Ranges for triggering
DPE events with Roman Pot trigger and rapidity gap trigger for the different
optics. Here, the symmetric case of equal momentum loss for the two protons
(i.e. $\xi_{1} = \xi_{2}$) is considered.}
\label{fig_diffaccept}   
\end{figure}

Depending on the running scenarios with different beam foci and 
luminosities, the acceptances for leading protons change considerably.
Fig.~\ref{fig_diffaccept} (left) shows the $(t, \xi)$ acceptances for
low, intermediate and high $\beta^{*}$ optics. 
For $\beta^{*} = 0.5\,$m, all protons with $\xi > 2\,\%$ are observed, 
almost independently of their $t$-values. For the two other optics, 
all $\xi$-values down to $10^{-8}$ are accepted for 
$|t| > 2\times 10^{-3}~(3\times 10^{-2})\,\rm GeV^{2}$ for
$\beta^{*} = $1540\,m (90\,m). Consequently, a large fraction of the 
diffractive protons is observed. In particular, the 90\,m optics will allow 
the study of semi-hard and hard diffraction, already at early LHC runs.

\begin{figure}[h!]
\begin{center}
\mbox{
\epsfig{file=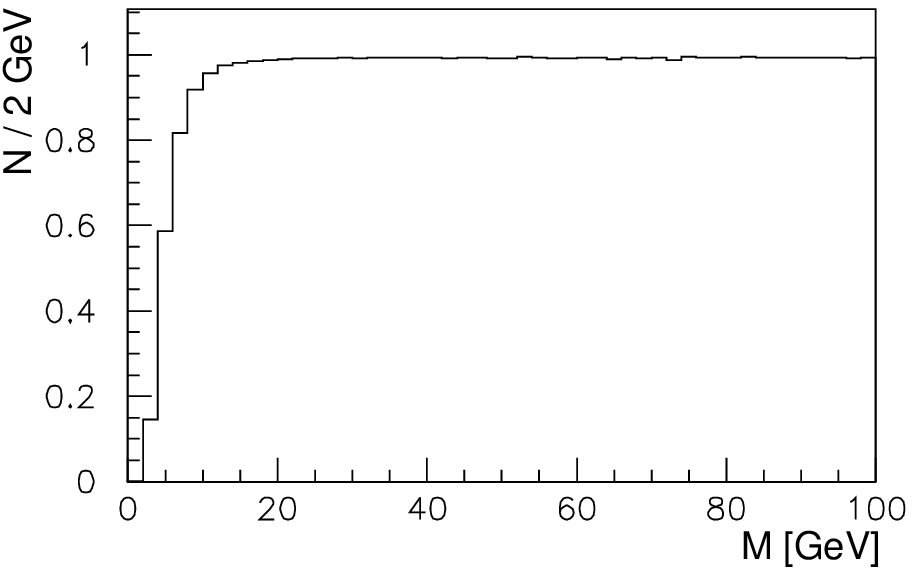,width=0.45\textwidth}
\epsfig{file=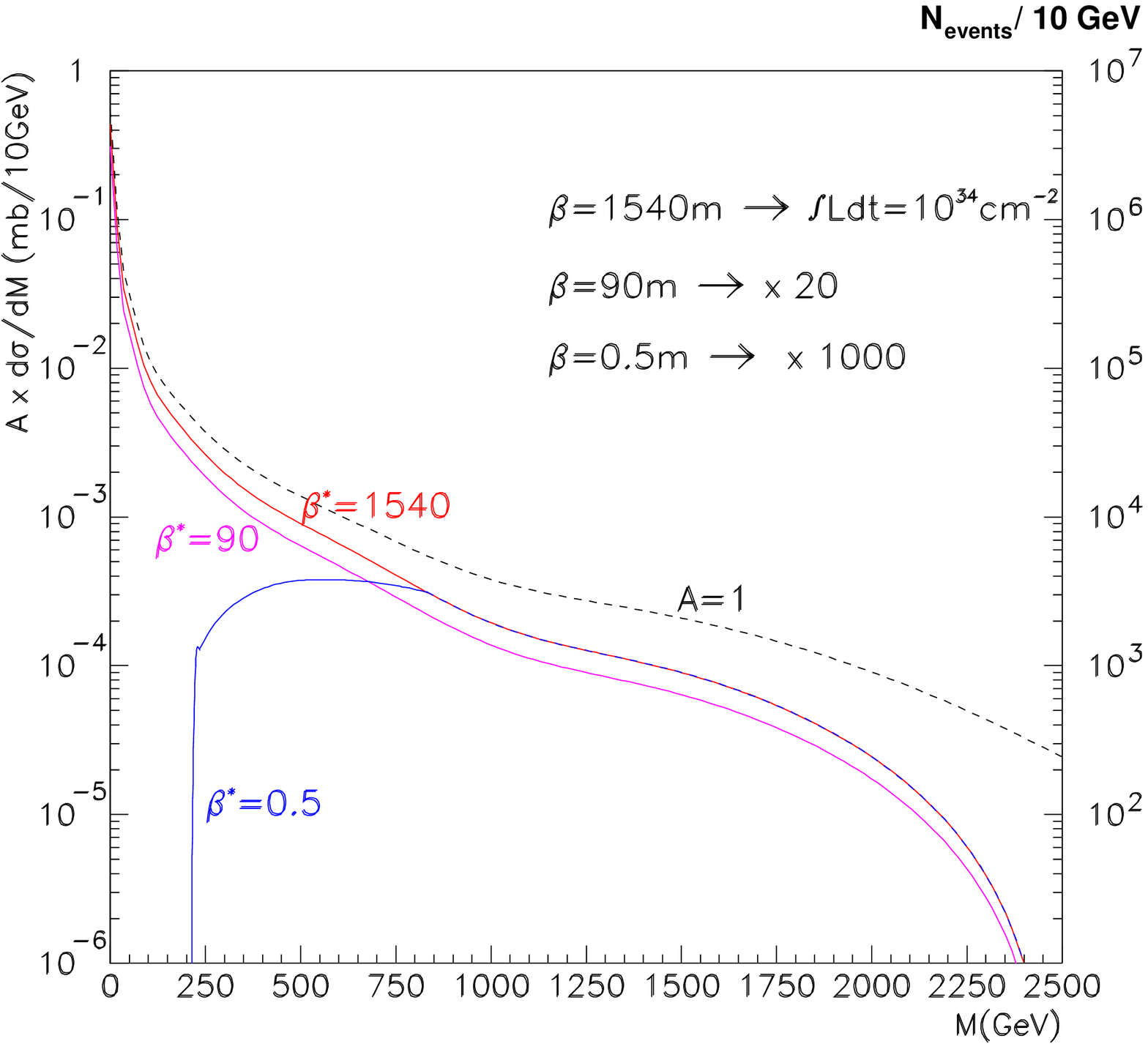,width=0.55\textwidth}
}
\end{center}
\caption{Left: acceptance for Single Diffractive states with mass $M$ using
the forward trackers T1 and T2. Right: differential cross-section of Double 
Pomeron Exchange without (dashed) and with (continuous lines) 
accounting for the leading-proton acceptance of the Roman Pots for different
beam optics.
}
\label{fig_massaccept}   
\end{figure}
Diffractive events are identified in the cleanest possible way by the 
observation of the forward protons.
However, with its large rapidity acceptance in the forward regions, TOTEM 
can also use the rapidity gap method (measuring the gap between the most 
forward particle 
and the proton on the same side) that allows triggering on low $\xi$ values 
even at $\beta^{*} = 0.5\,$m, thus complementing the direct proton measurement.
Fig.~\ref{fig_diffaccept} (right) shows the $\xi$ acceptance of the 
Roman Pot trigger and of the rapidity trigger. For $\beta^{*} = 0.5\,$m, 
the two trigger possibilities are just overlapping thus covering the complete 
$\xi$ range. How often the rapidity gap survives and is not filled by gluon 
radiation can be checked with the two other optics scenarios which offer 
redundancy of the Roman Pot and rapidity gap triggers. 
This so-called ``rapidity 
gap survival'' is an important quantity when comparing HERA and hadron 
collider data. Also given on the right-hand scale is the corresponding 
diffractive mass, produced in DPE for the symmetric case $\xi_{1} = \xi_{2}$. 

The DPE mass distribution obtained with a proton trigger on both sides 
is shown in Fig.~\ref{fig_massaccept} (right), also multiplied 
by the acceptances of the three optics scenarios.
With the high and intermediate $\beta^{*}$ optics, all masses, down to the 
lowest values, can be observed, whereas the low $\beta^{*}$ optics introduces 
a mass cut around 250\,GeV but extends the high-mass reach due to the 
higher luminosity. On the right-hand scale, the number of events for an 
integrated luminosity of $10^{34}\rm \,cm^{-2}\,s^{-1}$ demonstrates that 
events with masses above 2\,TeV can be observed even with a one-day run 
at $\beta^{*} = 1540\,$m. 

The mass acceptance for Single Diffractive states 
(Fig.~\ref{fig_massaccept}, left) extends to masses as low as 10\,GeV and 
hence allows the measurement of the Single Diffractive cross-section without 
too much extrapolation.


\section*{Acknowledgements}
It is a pleasure for me to acknowledge the contributions of V.~Avati and 
M.~Deile.

\end{document}